%% file: main.tex
\documentclass[a4paper,twoside]{article}

\usepackage{epsfig}
\usepackage{subcaption}
\usepackage{calc}
\usepackage{amssymb}
\usepackage{amstext}
\usepackage{amsmath}
\usepackage{amsthm}
\usepackage{multicol}
\usepackage{pslatex}
\usepackage{apalike}
\usepackage{paralist} 
\usepackage[utf8x]{inputenc}
\usepackage[colorlinks=true,linkcolor=blue,citecolor=blue]{hyperref}%
\usepackage{float}
\usepackage[textsize=tiny]{todonotes}
\usepackage[linesnumbered,ruled,vlined]{algorithm2e}
\SetKwInput{KwInput}{Input}
\SetKwInput{KwOutput}{Output}
\usepackage{booktabs}
\usepackage{listings}
\usepackage{xcolor}
\definecolor{codegreen}{rgb}{0,0.6,0}
\definecolor{codegray}{rgb}{0.5,0.5,0.5}
\definecolor{codepurple}{rgb}{0.58,0,0.82}
\definecolor{backcolour}{rgb}{0.95,0.95,0.92}
\lstdefinestyle{mystyle}{
    backgroundcolor=\color{backcolour},   
    commentstyle=\color{codegreen},
    keywordstyle=\color{magenta},
    numberstyle=\tiny\color{codegray},
    stringstyle=\color{codepurple},
    basicstyle=\ttfamily\footnotesize,
    breakatwhitespace=false,         
    breaklines=true,                 
    captionpos=b,                    
    keepspaces=true,                 
    numbers=left,                    
    numbersep=5pt,                  
    showspaces=false,                
    showstringspaces=false,
    showtabs=false,                  
    tabsize=2
}
\usepackage{SCITEPRESS} 

\begin{document}

\title{A Verifiable Multiparty Computation Solver for the Assignment Problem and Applications to Air Traffic Management}

\author{\authorname{Thomas Lor\"unser, Florian Wohner and Stephan Krenn\orcidAuthor{0000-0003-2835-9093}}
\affiliation{AIT Austrian Institute of Technology, Vienna, Austria}
\email{\{thomas.loruenser, florian.wohner, stephan.krenn\}@ait.ac.at}
}


\abstract{
The assignment problem is an essential problem in many application fields and frequently used to optimize resource usage.
The problem is well understood and various efficient algorithms exist to solve the problem.
However, it was unclear what practical performance could be achieved for privacy preserving implementations based on multiparty computation (MPC) by leveraging more efficient solution strategies than MPC based simplex solvers for linear programs.
We solve this question by implementing and comparing different optimized MPC algorithms to solve the assignment problem for reasonable problem sizes.
Our empirical approach revealed various insights to MPC based optimization and we measured a significant (50x) speedup compared to the known simplex based approach.
Furthermore, we also study the overhead introduced by making the results publicly verifiable by means of non-interactive zero-knowledge proofs.
By leveraging modern proof systems we also achieve significant speedup for proof and verification times compared to the previously proposed approaches as well as compact proof sizes.
}

\onecolumn \maketitle \normalsize \setcounter{footnote}{0} \vfill

\renewcommand{\hypertarget}[2]{#2}

\providecommand{\tightlist}{%
  \setlength{\itemsep}{0pt}\setlength{\parskip}{0pt}}

\input{macros}

\input{sections/intro}

\input{sections/lsap}

\input{sections/security}

\input{sections/verifiability}
\input{sections/conclusion}

\section*{\uppercase{Acknowledgements}}
This work has received funding from the European Union's Horizon 2020 research and innovation programme under grant agreement No 890456 (SlotMachine), No 871473 (Kraken) and No 830929 (CyberSec4Europe).

\bibliographystyle{apalike}
{\small
\bibliography{references}
}

%

\end{document}

%% file: macros.tex
\newcommand{\ZKSetup}{$\mathsf{ZKSetup}$}
\newcommand{\RegUser}{$\mathsf{RegUser}$}
\newcommand{\RegEqm}{$\mathsf{RegEqm}$}
\newcommand{\ITT}{$\mathsf{ITT}$}
\newcommand{\Match}{$\mathsf{Match}$}
\newcommand{\ComInput}{$\mathsf{ComInput}$}
\newcommand{\Input}{$\mathsf{Input}$}
\newcommand{\CkInput}{$\mathsf{CkInput}$}
\newcommand{\Compute}{$\mathsf{Compute}$}
\newcommand{\ZKProof}{$\mathsf{ZKProof}$}
\newcommand{\Reveal}{$\mathsf{Reveal}$}

\newcommand{\SessionInit}{$\mathsf{SessionInit}$}
\newcommand{\Clearing}{$\mathsf{Clearing}$}

%% file: sections/intro.tex
\section{Introduction}%
\label{sec:intro}
Efficient use of resources is of utmost importance for high competitiveness and low prices for consumers.
With the increasing degree of digitization and the ongoing trend towards cloudification, it becomes easier than ever before to achieve the goal of efficient resource usage also beyond company boundaries, e.g., in a sharing economy, where an optimal match between supply and demand has to be found.


One important task in such a scenario is described by the linear assignment problem, which deals with the question how to assign $n$ tasks to $n$ machines while minimizing the total costs, knowing the costs of assigning each task to each machine.
Assignment problems have been studied for multiple decades, and a variety of efficient algorithms solving such problems can be found in the literature.


However, when assigning resources among competitors, e.g, by the means of an auction, companies - and in particular potential competitors - might often have confidentiality concerns, as the individual costs per task might be sensitive and contain company secrets.
This might lead to hiding the true valuations \cite{AusubelM02}, or let companies not participate in an auction at all \cite{DBLP:conf/sigecom/SunderamP03}, e.g., because the leaked information might be used against them by competitors \cite{Moldovanu12}.


The classic approach in such a case would be to agree on a trusted party that collects all inputs from all participants, and locally computes the optimal assignment.
However, finding such a trusted authority might be difficult in many situations, e.g., in the case of competitors from different countries, or in case of a high frequency of such assignments close to real-time.

In this work we thus study solvers for the linear assignment problem, guided by the following main requirements:
\begin{enumerate}[i)]
\tightlist%
\item\label{req:decentralized}
No central authority shall be required in the entire process, i.e., all computations need to be carried out in a distributed fashion. 
In particular, all sensitive input data needs to be protected from unauthorized access by any involved entity.
\item\label{req:verifiability}
The output of the distributed computation shall be publicly verifiable (or at least by all participants), without requiring to trust any other entity in the system.
\item\label{req:efficiency}
Computations need to be sufficiently efficient and scalable to support a high frequency of executions close to real-time.
\end{enumerate}

\vspace{-1.3em}
\paragraph{Motivating example.}
For specificity, we explain the motivating use case for our research, which was developed in close collaboration with relevant stakeholders from the aviation industry as an important step towards practical deployment \cite{SGPL21,LSG21}.

Deviations from original flight plans due to variance and external events such as changing weather conditions are part of the day-to-day business at airports.
Therefore, optimization of starting and landing sequences across competing airlines could contribute to minimizing costs or delays on a large scale.
To do so, the current situation at an airport would ideally be continuously monitored and optimized, thereby considering airline priorities for most efficient operations.

A first system called User-Driven Prioritization Process (UDPP) \cite{udpp} has already been developed to allow airlines to react on varying conditions and swap flights within their own fleet for given flight sequence.
However, because airlines are reluctant to share their preferences with other airlines, global optimization is currently not possible, which significantly reduces the efficiency of the given resources at an airport.

In this work we aim at tackling this problem by leveraging multi-party computation to develop a decentralized platform that enables collaboration for optimal flight sequencing in challenging conditions, cf. (\ref{req:decentralized}).
Based on dedicated market mechanisms, set up to incentivice airlines to participate in the system, a model for an optimization process was developed.
From a modelling point of view, a weight map is used by airlines to define flight priorities for particular slots in the flight sequence.
Looking at the modelling of the optimization problem, it turns out that it basically resembles a so called linear sum assignment problem (LSAP).

Furthermore, related to (\ref{req:verifiability}), given the financial and economic impact of slot assignments, airlines have a strong requirement regarding the authenticity of any slot assignment in order to overcome the risk of unjustified prioritization of a single airline.
Finally, slots need to be assigned multiple times per hour due to the high traffic volume at major airports, and the frequency of delays, changing weather conditions, etc., thus requiring computations to be carried out in seconds to minutes at most, cf.~(\ref{req:efficiency}).

\vspace{-1.3em}
\paragraph{Related Work}
In the following we provide a brief overview over related work.

Numerous privacy-preserving algorithms for different types of matching algorithms have been proposed in the literature.

Considering general two-party linear programming (LP), \cite{DBLP:conf/colcom/0001A06} provide an efficient protocol for semi-honest parties, as well as extensions to prevent certain malicious behaviour.
Linear programming using MPC was considered by \cite{Toft09,DBLP:conf/sac/Vaidya09,DBLP:journals/corr/HongVRL16,DDN+15}.

Regarding specific assignment tasks, \cite{DBLP:conf/fc/Golle06,DBLP:conf/ctrsa/FranklinGM07} provide a privacy-preserving version of the famous matching algorithm by \cite{journals/tamm/GaleS13}, based on mix networks and homomorphic encryption, however only considering a weak (passive) adversary model.
A first MPC-based implementation was presented by \cite{DBLP:conf/ccs/DoernerES16}, scaling to multiple thousand input values. 
A first provably secure and scalable implementation was later presented by \cite{DBLP:journals/popets/RiaziSSSK17} based on garbled circuits~\cite{DBLP:conf/focs/Yao86}.

Specifically for LSAP, a privacy-preserving version of the Hungarian algorithm (as in Section~\ref{ssec:munkres}) based on homomorphic encryption was presented by \cite{Wuller2017a}.
However, only the theoretical complexity of the protocol is analyzed, and no performance data is available.

None of these protocols offers means to publicly verify the correctness of the computation result, which however is a key requirement for our use case.
A notable exception is the work by \cite{HSM2016}, who present verifiable MPC-based solutions for general linear programming.
However, due to the generality of LP as well as the choice of primitives for the correctness proofs, our efficiency requirements cannot be achieved by this work.
An efficient publicly verifiable auctioning platform for traditional sealed-bid auctions was recently proposed by \cite{DBLP:conf/icissp/LorunserWK22}.

For the sake of completeness, we also mention \cite{DBLP:conf/scn/BaumDO14,SM2015} who give a generic framework for publicly verifiable multi-party computation, which however is mainly of theoretical interest in our setting due to the computational overhead.

\vspace{-1.3em}
\paragraph{Contributions.}
  Following the above guiding principles, the main contributions of this paper can be summarized as follows:
\begin{itemize}%
  \tightlist%
  \item 
	  In a first step, we perform a comprehensive analysis and comparison of secure multi-party computation (MPC) based approaches to solve the assignment problem in a privacy-preserving way.
  \item 
	  We provide optimized implementations and benchmarks to compare the performance of different approaches, achieving an improvement over existing implementations by a factor of $50$.
  \item 
	  We extend our implementation by public verifiability mechanisms based on zkSNARKs and Bulletproofs, thereby significantly outperforming related work and demonstrating the practical efficiency of decentralized, privacy-preserving, verifiable solvers for the assignment problem.
\end{itemize}

\vspace{-1.3em}
\paragraph{Outline.}
The remainder of the paper is structured as follows.
We provide a comprehensive overview of MPC-based approaches to the assignment problem in Sec.~\ref{sec:lsap}.
In Sec.~\ref{sec:evaluation}, we present our implementation approach and practical benchmark results to solve the assignment problem, and present our extension towards public verifiability in Sec.~\ref{sec:verifiability}.
We briefly conclude in Sec.~\ref{sec:conclusion}.

%% file: sections/lsap.tex
\section{MPC Approaches to LSAP}%
\label{sec:lsap}

In this section we briefly review the linear sum assignment problem (LSAP) and discuss important aspects when it comes to the realization of a privacy-preserving version based on MPC.

\subsection{Assignment Problem}

An instance of LSAP is described by a weight matrix $W$, where each $w_{i,j}$ represents the cost associated with matching task $i$ of the first set (a flight in our case) and resource $j$ of the second set (a slot in our case).
The goal of the optimization is then to find a complete assignment of flights to slots which is of minimal cost according to a defined objective function, which is essentially the sum of weights.

Formally, let $X$ be a boolean matrix where $x_{ij} = 1$ if and only if row $i$ is assigned to column $j$.
Then the cost of the optimal assignment is computed as

$$
\min \sum_i \sum_j w_{ij} x_{ij}\,
$$

where the minimum is taken over all $X$ where each row is assignment to at most one column, and each column to at most one row.
In our analysis the matrix $W$ was assumed to be quadratic, however, it can be easily generalized to a rectangular problem by techniques discussed below.

A large number of algorithms has been developed for the LSAP, cf., e.g., \cite{Akgul1992,Bertsekas1992,Burkard1999,DT00}. 
They range from primal-dual combinatorial algorithms, to simplex-like methods, cost operation algorithms, forest algorithms, and relaxation approaches.
The worst-case complexity of the best sequential algorithms for the LSAP is $O(n^3)$, where $n$ is the size of the problem.

For this work we selected one representative for each important class of algorithms and analyzed/implemented a MPC version of it to measure the practical performance which can be achieved.
The selected algorithms are the
\begin{itemize}
\tightlist%
\item simplex based solution strategy, where we leveraged linear programming to converted the problem into max flow formulation.
\item Hungarian algorithm (aka Munkres), one of the most important candidates for the primal-dual strategy,
\item auction algorithm, a algorithm working in the dual domain of "shadow prices", and
\item variants of shortest augmenting path (SAP) algorithms.
\end{itemize}




If the weight matrix is quadratic in size $n$, i.e., there is the same number of tasks and resources, the LSAP is called \emph{balanced}.
It means that both parts of the bipartite graph have the same number of vertices, when treating the problem as matching in bipartite graphs.

In the \emph{unbalanced} case, the number of vertices is different for each side in the corresponding bipartite graph, resulting in a rectangular cost matrix $n \times m$.
In that case, either not every machine can be matched to a task or not every task is occupied.
Fortunately, most of the algorithms tested can be directly generalized to unbalanced problem solving.
However, even if the solver only works for balanced problems, there are methods to convert an unbalanced solution to a balanced one.
The straight forward technique is to augment the smaller set of vertices with $|n-m|$ additional entries and to connect them to the existing vertices with edges of cost 0.
However, there also exist even more efficient technique \cite{doubling-technique} requiring even less additional edges. 
Fortunately, all this techniques are also compatible with MPC and only result in an additional pre-processing step.

\subsection{MPC Aspects}
Multi-party computation (MPC) allows parties to jointly perform computations in a way that only designated receivers obtain a result at the end of the computation, while no further information is revealed to any other participant in the system.
In particular, the inputs are kept confidential from all other participants in the system.
MPC can be considered the most practical approach for generic computation on sensitive data. 
It allows to perform arbitrary computations in principle, however, depending on the concrete computation to be performed, MPC protocols are often slower than a local computation by orders of magnitudes.

Generally speaking, the algorithms used to solve the LSAP are not MPC-friendly.
By their nature, they are mostly sequential with very little potential for vectorized operations.
One such vectorizable operation is testing for zero.
Even though this is a costly procedure in MPC that involves random number generation and comparisons, it can easily be done for a whole array in parallel, because testing one element does not involve any other elements of the same array.
Also, the result can be cached, is only invalidated if the value itself changes, and can easily be recomputed on demand.

With most other operations, however, this is not possible.
Take for example the minimum of a collection of elements.
Finding it involves in the order of log n comparisons that have to be performed in sequence.
Any change of the collection over which the minimum was computed could possibly change the minimum, so caching it is not viable.
(When an element is added or changed, a single comparison is sufficient to recompute the minimum, but when an element is removed, the minimum has to be recomputed from scratch.)

To get tolerable performance we must trade-off between privacy and speed and inevitably leak some indirect information, e.g, branches been taken. However, the final assignment will be public and is known to be optimal, which also means some leakage. 
If that is not enough, \cite{AC2017} have shown how to efficiently implement graph algorithms that, like ours, reveal branching information, yet do not leak information
by just obliviously permuting the original data.

Another problem is that every algorithm that uses some form of $\epsilon$-scaling needs to use floating-point numbers.
This is not just a question of numerical stability.
If the underlying numerical representation is not precise enough, $\epsilon$-scaling may terminate with a solution that is not optimal, or may not even terminate at all.
In~\cite{Bertsekas2009} the authors propose to multiply every element of the $n*n$ matrix by $(n+1)$ and use only integer values (down to 1) for $\epsilon$ but notes that this may in practice lead to integer overflow because prices can then be somewhere in the order of $n^2 \max_{(i,j) \in A} | a_{ij} |$.

\section{Algorithm Evaluation}%
\label{sec:evaluation}

In the following we compare MPC performance of different solution strategies used to solve the assignment problem.
The different algorithms have been implemented and benchmarked in \textit{MPyC}\footnote{\url{https://github.com/lschoe/mpyc}} with default settings and a 3 party configuration.
\textit{MPyC} is based on secret sharing and is targeted towards semi-honest adversaries, however, the results can also be transferred to other frameworks with reasonable effort.
The performance of the simplex solver from~\cite{HSM2016} served as a baseline for our comparison and was included in our analysis as shown below.
A single Intel NUC computer equipped with an Intel(R) Core(TM) i5-8259U CPU running at 2.30GHz maximum frequency and with 32GB of memory was used as hardware, to make the results comparable.
All parties were run in a local setup without any additional network latency and other restricting settings, if not explicitly stated otherwise.
If not explicitly stated otherwise, all presented runtime are in seconds.

\subsection{Simplex for Linear Programming}

The assignment can be viewed in different forms. 
In essence, it is a special case of the transportation problem, which itself is a special case of the minimum cost flow problem, which belongs to category of linear programs.
Therefore, the most generic solving approach would be to leverage existing simplex implementations in MPC and model the problem accordingly.

\begin{figure}[ht]
  \centering
  {\epsfig{file=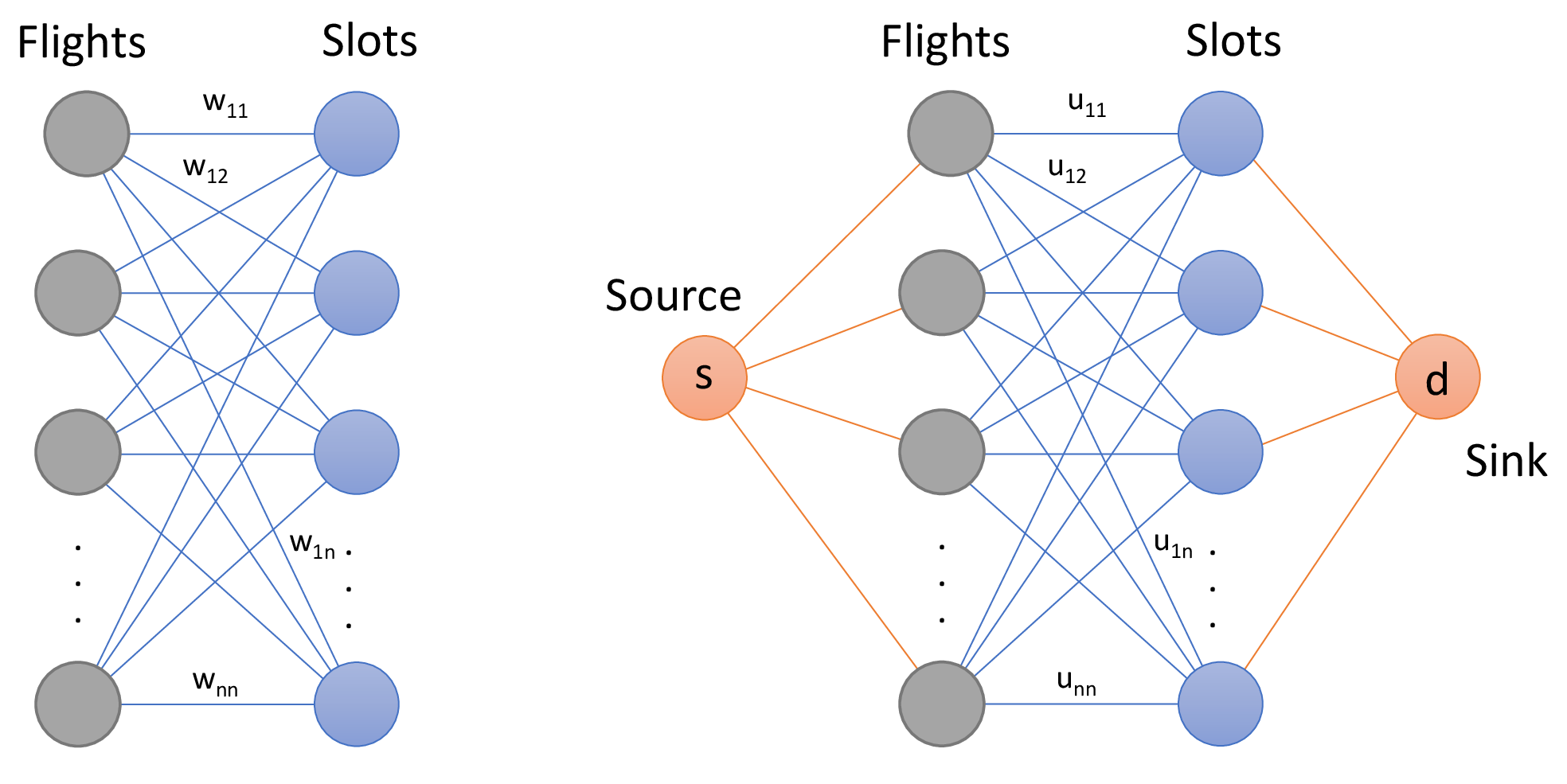, width=0.5\textwidth}}
  \caption{Bipartite graph structure for matching and minimum cost flow modelling.}%
  \label{fig:lpmodel}
\end{figure}

The two major representations in LP form are shown in Fig.~\ref{fig:lpmodel}.
They are either modeled as minimum cost matching in a bipartite graph or min cost flow problem.
The latter may be rather counter intuitive because one would more likely expect the formulation as an integer program because of the binary nature of a match.

The corresponding LP is defined as follows.
In a bipartite graph each edge $(i,j)$, where $i$ is in $A$ and $j$ is in $T$, is assigned a weight $w_{ij}$.
Additionally, for each edge $(i,j)$ we have a binary variable $x_{ij}$ indicating if a certain edge is in the solution or not.
%
%
Therefore, the resulting LP is given by:
\begin{equation}%
\label{eqn:lp}
\begin{array}{l l}
  \text{minimize} & \sum_{(i,j)\in A\times T} w_{ij}x_{ij} \\[1em]
  \text{subject to} & \sum_{j\in T}x_{ij}=1\text{ for }i\in A, \, \\
                    & \sum_{i\in A}x_{ij}=1\text{ for }j\in T \\[1em]
   \text{with}      & 0\le x_{ij}\le 1\text{ for }i,j\in A,T, \, \\
                    & x_{ij}\in \mathbb{Z}\text{ for }i,j\in A,T.
\end{array}
\end{equation}

Because of the binary variables the model resembles an integer linear program.
Fortunately, the problem can still be solved with standard methods known from continuous LP, albeit the integrality constraints, by simply dropping the integrality constraint. 
This is due to the fact, that for optimal solutions variables always take integer values, despite fractional values being allowed.

Furthermore, converting the problem to a maximization solution by inverting the weights leads to a further simplified formulation with less slack variables.
In the presented use case we where anyhow maximizing the utility which is represented by the cost.
Reducing the number of slack variables and problem size is essential for MPC performance and by converting the equality constraints to $\sum_{j\in T}x_{ij} \leq 1$ and $\sum_{i\in A}x_{ij} \leq 1$ the most compact formulation is achieved.

\begin{table}[ht]
\centering
\begin{tabular}{cccccc}
\toprule
$s$ & $n$ & $m$ & $iter$ & $t_{simplex}$ & $t_{dual}$ \tabularnewline
\midrule
10& 100	& 20 & 23	& 7.8 & 1.1 \\
20& 400	& 40 & 41	& 61.8& 12.8\\
30& 900	& 60 & 71	& 275 & 57  \\
40& 1600&	80 & 104& 806 & 157 \\
50& 2500& 100& 145& 1920& 410 \\
60& 3600& 120& 167& 3468& -   \\
70& 4900& 140& 224& 7333& -   \\
\midrule
10&	100	&  20& 21	& 7.1	& 1.1 \\
20&	400	&  40& 41	& 59.7& 11.9\\
30&	900	&  60& 66	& 253	& 57  \\
40&	1600&  80& 86	& 690	& 180 \\
50&	2500&	100& 117&	1643&	378\\
60&	3600&	120& 140&	2869&	-  \\
\bottomrule
\end{tabular}
\caption{Running time in seconds of LP solver with random weight vectors (above) and sample data from slot management problem (below). $s$ is the size of the quadratic weight matrix and $n \times m$ is the dimension of the respective dimension of the $A$ matrix in a LP of the form $Ax \leq b$.}%
\label{tab:lpbench}
\end{table}

The implementation used is based on the simplex version presented in \cite{HSM2016}
and the results of our performance measurements are shown in Table~\ref{tab:lpbench}.
The upper part in the table are benchmarks for randomly generated weight matrices and the lower part is for typical sample data from our use case.
Because the use case data is more structured slightly better runtimes can be expected, but the improvement is not significant.
Even worse, the implementation was not able to generate the dual certificate also incorporated in the implementation because of the high memory usage required for problem sized bigger than $50$.
Furthermore, also the implementation itself also stopped working because of networking problems for problem sizes beyond $70$.
We did not further investigate this behaviour as we were interested in alternative solution approaches anyway, however, this measurement served as a reference for our other implementations.

\subsection{Hungarian Method}\label{ssec:munkres}

Our second implementation is based on the Hungarian algorithm, also known as the Munkres or Kuhn-Munkres algorithm \cite{Kuhn1955,Kuhn1956,Munkres1957}.
It was one of the first polynomial-time algorithms for solving the assignment problem.
The basic idea of Munkres algorithm is to iteratively improve the matching in a bipartite graph along augmenting path between unmatched vertices.
It has the fastest strongly polynomial run-time complexity with $O(mn+n^{2}\log n)$, where $n$ is the number of vertices and $m$ is a number of edges, when implemented with Fibonacci heaps.



Our MPC version is based on a standard implementation as presented in \cite{Toft09}.
Contrary to the original algorithm for manual evaluation with 4 phases, it comprises 6 steps but follows the main paradigm of finding minimum coverings of zeros in the weight matrix manipulated by reducing rows and columns.

However, a fully oblivious implementation would be rather slow and would require further measures to prevent from leaking information.
Therefore, we opted to reveal certain aspects during computation, which will be discussed in Sec.~\ref{s:leakage}.
With this approach we achieved a substantial performance speedup compared to the simplex variant and after some manual optimization we achieved a speedup of almost a factor of $30$ compared to the simplex.

\begin{table}[ht]
\centering
\begin{tabular}{ccccc}
\toprule
size & \#steps & $t_{munkres}$ & \#iszero & \#min \tabularnewline
\midrule
10  &  38   & 0.8  & 400    & 821    \\
20  &  86   & 4.4  & 1981   & 4016   \\
30  &  184  & 17.8 & 10309  & 20734  \\
40  &  308  & 33.7 & 26368  & 52931  \\
50  &  441  & 67.0 & 52714  & 105712 \\
60  &  601  & 125  & 98444  & 197280 \\
70  &  775  & 170  & 154046 & 308603 \\
80  &  962  & 220  & 230023 & 460682 \\
90  &  1193 & 345  & 336901 & 674583 \\
100 &  1401 & 412  & 460890 & 922709 \\
\midrule
10& 36& 0.9& 399& 818\\
20& 92& 5.0& 2633& 5323\\
30& 185& 16.1& 10219& 20556\\
40& 341& 53.0& 34020& 68253\\
50& 488& 77.8& 67730& 135772\\
60& 725& 215& 149425& 299310\\
70& 959& 293& 264503& 529618\\
80& 1183& 350& 396112& 792990\\
90& 1450& 546& 591673& 1184293\\
100& 1770& 869& 884127& 1769411\\
\bottomrule
\end{tabular}
\caption{Running time in seconds of Munkres with random weight vectors (above) and sample data from slot management problem (below).}%
\label{tab:munkresbench}
\end{table}

The detailed performance results are shown in Table~\ref{tab:munkresbench}.
The table shows the duration of an optimization run in seconds depending on the problem size.
It also contains information about the amount of costly MPC operations needed (minimum finding and zero testing) in the processing.
Interestingly, for Munkres the average performance measured for the random case is twice as fast as for the particular use case data which is more structured.
This is in contrast to the simplex solver where the algorithm could benefit from the structure in the use case data.

\subsection{$\epsilon$-scaling Auction Algorithm}

The auction algorithm is an intuitive method for solving the classical assignment problem.
It was first introduced in 1979 by \cite{Bertsekas1979}, and has since then evolved as a valuable tool in network optimization \cite{Bertsekas2009}.
Auction algorithms were selected for implementation because they have good practical average performance, although worst case performance is the same as for the Hungarian algorithm, i.e., $O(n^3)$.


In this paragraph we quickly recap the description from \cite{Bertsekas2009}, which is based on the idea of economic equilibrium problem that turns out to be equivalent to the assignment problem;
for a detailed presentation, we refer to \cite{Bertsekas1992,Boffey1994,Bertsekas1998}. 

The auction algorithm works in the dual of the problem acting on the so called \emph{shadow prices}.
In a first step, it determines $\epsilon$ to be the highest absolute cost, then repeats the auction with progressively smaller $\epsilon$ until it is smaller than $n$ (the number of participants/objects).
The rate $\alpha$ of decrease can be freely chosen.

This seems to indicate that floating point numbers have to be used in order to guarantee correctness and termination of the algorithm.
However, as mentioned before, it is possible to scale all costs by $n+1$ and remain in the integer domain, with the risk of integer overflows.

In practice, the auction algorithm is very MPC-unfriendly.
To find the initial $\epsilon$ involves taking the maximum of the whole cost matrix, and afterwards repeatedly finding the two indices at which the current price vector is minimal.
As the price vector is highly variable, there is no possibility of caching.
Given the high overhead of floating-point arithmetic in our development environment, initial benchmarks showed a slowdown of a factor of more than $10$ compared to all other solutions already for small problem sizes, such that this type of algorithms was not further considered in our analysis.

\subsection{Shortest Augmenting Path Algorithms}

Another important category of algorithms are shortest augmenting path algorithms
such as the Jonker-Volgenant-Castanon (JVC) \cite{Jonker1987}.
These algorithms are somewhat similar to Hungarian method, but apply a better way to update solutions together with a number of pre-processing techniques, including column reduction, reduction transfer, and reduction of unassigned rows.
While the Hungarian algorithm finds any feasible augmenting path, JVC and a number of other algorithms find the shortest augmenting paths in a minimum cost network flow, where each node in $S$ transmits one unit and each unit in $T$ must receive one unit of a single commodity.
Indeed, an optimal solution can be found by considering one source in $S$ at a time and finding the shortest path emanating from it to an unassigned node in $T$.


We compared two implementations of this class.
The first implementation is based on a solution used in the optimization module of SciPy module%
\footnote{\href{https://github.com/scipy/scipy/blob/v1.7.0/scipy/optimize/rectangular\_lsap/rectangular\_lsap.cpp}{Github scipy package file \textit{rectangular\_lsap.cpp}}}.
The second implementation is based on the py-lapsolver project\footnote{\url{https://github.com/cheind/py-lapsolver}}, which itself is based on the Stanford ACM-ICPC teams site%
\footnote{\href{https://github.com/jaehyunp/stanfordacm/blob/master/code/MinCostMatching.cc}{Github StandfordACM algorithm \textit{MinCostMatching.cc}}}.

\begin{table*}[ht]
\centering
\begin{tabular}{ccccccccccc}
\toprule
 & random & use case &  & & random & use case && \tabularnewline
n & $t_{scipy}$ & $t_{scipy}$ & \#iszero & \#min
  & $t_{lapsolve}$ & $t_{lapsolve}$ & \#iszero & \#min \tabularnewline
\midrule
10	& 1.2	  & 1.7	 & 641	  & 199   & 0.7	 & 0.9	 & 492	  & 380   \\
20	& 7.1	  & 11.9 & 5155	  & 998   & 3.4	 & 4.5	 & 2860	  & 1940  \\
30	& 20.4  & 35.3 & 17535	& 2797  & 8.2	 & 10.4 & 8322	  & 5480  \\
40	& 35.9  & 64.3 & 42641	& 5996  & 17.0 & 22.9 & 18429  & 11800 \\
50	& 74.0  & 97.5 & 78803	& 10995 & 34.9 & 38.2 & 34280  & 21700 \\
60	& 128	  & 135	 & 126972	& 18194 & 56.5 & 57.4 & 58762  & 35980 \\
70	& 176	  & 174	 & 186655	& 27993 & 86.7 & 73.5 & 88404  & 55440 \\
80	& 262	  & 224	 & 267247	& 40792 & 116  & 103	 & 129582 &	80880 \\
90	& 393	  & 255	 & 381406	& 56991 & 149  & 138	 & 182114 &	113100\\
100	& 438	  & 289	 & 526712	& 76990 & 185  & 188	 & 247934 &	152900\\
\bottomrule
\end{tabular}
\caption{Running time in seconds of scipy\_lsa and lap\_solver with random weight vectors and sample data from slot management problem. Measurement of number of zero tests and minimum search were done on use case data.}%
\label{tab:sapbench}
\end{table*}

The performance results of the MPC implementation are summarized in Table~\ref{tab:sapbench}.
For random data the SciPy version performs similar to Munkres, but the algorithm also benefits from the structure in typical use case data.
However, the MPC version of the ACM-ICPC solver turned out to be the fastest in class and also in general.
It is more than two times faster than the SciPy version and more than four times faster than Munkres for the use case specifc data.
Nevertheless, in this version the smart updating mechanism based on augmenting paths did not make a real difference for different data types.
Overall, the MPC-ACM-ICPC solver performs about $55$ times better than the state of the art based on simplex solver which pushes practical applications to bigger problem sizes as in the case of the markets for air traffic management slot exchange.



\subsection{Impact of Network Latency}
To make the results comparable all benchmarks were done on the same single PC with the very same software framework and the same configuration.
However, no delays between network nodes have been introduced.
In principle network delays increase the time during sub-protocols for non-linear operations, i.e., multiplication steps in our protocols (c.f. see \cite{LRW22}).
Therefore, the experienced slowdown depends linearly on the multiplicative depth of the arithmetic circuit defining the function to be computed.
To show the impact in practical terms, in Table~\ref{tab:lsap-latency} we show the impact of latency between MPC nodes.
Real world solutions have to take this effect into account.
It could lead to substantial performance penalties for distributed setups with larger network latency \cite{LW20}.

\begin{table}[ht]
\centering
\begin{tabular}{cccccc}
\toprule%
\hspace{1.6em}n → & 10 & 20 & 30 & 40 & 50\tabularnewline%
latency ↓ &&&&\tabularnewline%
\midrule
0	 & 0.7  & 3.5	  & 9.1	  & 16.4	& 34.8 \\
5	 & 5.3  & 30.4	& 72.6	& 134.3	& 305  \\
10 & 10.1	& 56.4	& 132	  & 239	  & 536  \\
15 & 14.4	& 80.2	& 184	  & 335	  & 751  \\
20 & 18.4	& 102	  & 238	  & 435	  & 978  \\
\bottomrule
\end{tabular}
\caption{Performance of MPC version of ACM-ICPC solver for different network delays and problem sizes. Latency is in milliseconds.}%
\label{tab:lsap-latency}
\end{table}

%% file: sections/security.tex
\subsection{Leakage and Countermeasures}\label{s:leakage}


To achieve a good performance, trade-off between privacy and efficiency had to be accepted, which we will discuss in the following.

The Hungarian method is a completely sequential algorithm, and branching encodes information.
Therefore, a fully oblivious version would run in constant time and not even reveal the number of iterations needed.
However, this is not practical and render the technology obsolete for the aspired goal and problem sizes.
Certain trade-offs were already considered in~\cite{HSM2016}, where also minimal information is leaked by the algorithm to achieve better performance, e.g., the number of iterations.


In our implementation we never reveal costs at any time.
Yet, we carry out certain tasks, such as the row and column covering, in the plaintext domain.
This reveals information about the position of minimum elements in rows and column by doing public zero testing after the minimum of certain rows and columns have been subtracted obliviously.
This leakage could enable an observer (e.g., a semi-honest MPC node) to learn certain aspects about the structure of the cost matrix by following the covering results for rows and columns over the iterations.

Similarly, for the class of shortest augmenting path algorithms, the MPC-ACM-ICPC implementation is completely sequential and therefore leaks in case of non-oblivious branching, while never revealing costs themselves.

To cope with certain leakages, dedicated countermeasures can be put in place in form of pre- an post-processing.
For instance, for the Hungarian method, the leakage of the index of the minimum cost in a row or column can be defeated by obliviously permuting rows and columns before running the algorithm and reversing the operation after a solution is found.
This is due to the fact, that an optimal solution is given by a full covering will all columns marked, i.e., reversing the permutation fully removes all information about the intermediate steps for an adversary.
Computing permutations can be done highly efficiently with only a minimal overhead compared to the actual optimization process.
This approach still enables reasonable performance but prevents from attributing row an column properties to the real cost matrix.

To be more concrete and estimate the overall runtime of the pre- and post-processing phase we implemented an oblivious shuffle2d and unshuffle2d algorithm permuting rows and columns of an $n \times n$ cost matrix through appropriate matrix multiplications.
The runtimes are shown in Table~\ref{tab:shuffle}.

\begin{table}[H]
\centering
\begin{tabular}[]{@{}ccc@{}}
\toprule
n  & shuffle2d (s) & unshuffle2d (s) \tabularnewline
\midrule
10 & 0.1 & 0.02\\ 
50 & 1.1 & 0.2\\ 
100& 5.5 & 0.9\\ 
\bottomrule
\end{tabular}
\caption{Performance of permutation based pre- and post-processing.}%
\label{tab:shuffle}
\end{table}


%% file: sections/verifiability.tex
\section{Public Verifiability}%
\label{sec:verifiability}
As discussed earlier, a solver for the assignment problem should not only protect the privacy of the inputs, but also give formal guarantees about the correctness of the result.
That is, the computation result should also come with a cryptographic certificate (or proof) that allows any party to check whether all computations have been carried out correctly, without leaking any information about the inputs.
Such an approach, called \emph{publicly verifiable MPC}, minimizes the trust that needs to be put into the MPC network, as soundness can even be guaranteed in case that all MPC nodes get corrupted.
To achieve this, we deploy non-interactive zero-knowledge proofs of knowledge (NIZK)~\cite{DBLP:conf/stoc/BlumFM88}.



The idea for this type of optimization was introduced in \cite{HSM2016}.
The approach is based on the duality theorem which defines for every linear program an equivalent problem in the dual with dual variables and a dual objective function.
In fact, the efficient algorithms presented before already make use of the duality internally, which make them also good candidates for extension with verifiability.
By directly proving the optimality of the dual solution, it is no longer necessary to prove the correctness of every single computational step, which would not be feasible in a reasonable amount of time.

The dual of the linear assignment problem in (\ref{eqn:lp}) is as follows:
\begin{equation}\label{eqn:dual-lsap}
\begin{array}{l l}
  \text{maximize} & \sum u_i + \sum v_j \\[1em]
  \text{subject to} & u_i + v_j \leq c_{ij} \text{ for } i,j \in A,T  \, \\
   \text{with}      & u_i, v_j \in \mathbb{Z}\text{ for }i,j\in A,T.
\end{array}
\end{equation}%
where $u_i$ and $v_i$ are the dual variables which an not restricted to positive values contrary to the primal variables.
The dual variables are also interpreted as shadow prices which is why operations in the dual are often called auctions.

Thus, in order to efficiently prove the correctness of a optimization result we first need to compute the corresponding dual, which has to be kept private.
This can be achieved by augmenting the MPC algorithms to also compute the dual in the oblivious domain and only reveal the primal solution in the clear.
Secondly, in order to prove optimality, it is necessary to show that:
\begin{enumerate}
  \tightlist
  \item the optimum is indeed the sum of the costs (NIZK or directly),
  \item the constraints of the primal are fulfilled (this can be done in clear),
  \item there exist dual variables such that the primal optimum is equal to the dual, i.e. $\sum u + \sum v = f_{opt}$ (NIZK), and
  \item the dual variables fulfil the constraints (NIZK).
\end{enumerate}

The most challenging task are the last two steps which have to be done without revealing the dual variables or the costs which requires the usage of NIZK and the generation of them within the MPC system, without revealing any witness to any entity in the clear.
In the following we explain how this can be achieved efficiently using MPC and NIZK.


\subsection{Augmented LSAP}
In a first step, the optimization algorithms have to be adapted to also provide the dual solution.
We use the idea of augmented algorithms which provide both, an optimal solution of the original problem as well as the corresponding dual variables.

\vspace{-1em}
\paragraph{Verifiability for general LP.}
For the case of LP the approach has already been demonstrated in \cite{HSM2016}, and we use this implementation as a baseline for our improvements.
The respective solution also comes with verifiability, and consists of four main steps:
\begin{enumerate}
  \tightlist
	\item the basic simplex operation, 
	\item computation of the solution, 
	\item computation of the dual, and 
	\item verification of the dual and optimality.
\end{enumerate}
Unfortunately, especially the verification step turned out to be very resource intensive.
Besides adding another 20\% overhead to the computation time, especially the memory usage was extensive.
As shown in Table~\ref{tab:lpbench}, we were not able to conduct tests beyond problem sizes of 40 slots on our test machine, rendering the implementation impractical for our requirements.

\vspace{-1em}
\paragraph{Overcoming limitations for LSAP.}
In our work we thus developed and tested an augmented version of the Hungarian algorithm, as it also follows a primal-dual approach.

The core extensions to the Hungarian algorithm are shown in Listing 1 and 2.
In step 3 of the algorithm the $u_i$ have to be updated with each row modification
and in step 6 $u_i$ and $v_j$ are updated according row and column modifications.
All other steps are not affected and the necessary modifications in steps 3 and 6 are also very MPC friendly, i.e. only addition and subtraction on secure values.

\begin{lstlisting}[language=Python, caption=Augmented Munkres Step 3]]
for i in range(n):
    # Find min value for each row
    minval = min(cost_matrix[i])
    # Subtract minval from every element in the row.
    for j in range(n):
        self.C[i][j] -= minval
    # Update dual u_i 
    self.u[i] = minval
\end{lstlisting}

\begin{lstlisting}[language=Python, caption=Augmented Munkres Step 6]
for i in range(self.n):
    for j in range(self.n):
        if self.row_covered[i]:
            self.C[i][j] += minval
            if j == 0: self.u[i] -= minval
            events += 1
        if not self.col_covered[j]:
            self.C[i][j] -= minval
            if i == 0: self.v[j] += minval
            events += 1
        if self.row_covered[i] and not self.col_covered[j]:
            events -= 2 # change reversed, no real difference
\end{lstlisting}

%
%
Because shortest augmenting path (SAP) algorithms are very similar in nature to Hungarian, we expect similar results for them although we did not implement them.

\subsection{Adaptive zkSNARKS}
We now explain how the necessary NIZKs to prove the optimality of the dual solution are computed in our system, thereby significantly improving over the efficiency achieved by \cite{HSM2016}.

In our first approach we use the adaptive zkSNARKs by \cite{DBLP:conf/africacrypt/Veeningen17}, which are well suited and optimal in terms of proof size.
The idea is to have commitments on all relevant witnesses for the proof, which could, e.g., be stored on a blockchain to make them publicly available.

As described above, the proof is composed of four components where three have to be shown in zero-knowledge and the evident plaintext constraints are omitted.
Because the commitment used in the zkSNARKs system are homomorphic, the correctness of the optimum can be shown directly by combining the respective commitments on the input weights.
However, if the more efficient vector commitments are used a dedicated proof has to be computed explicitely in the MPC system, which is also straight forward.
The same has to be done for the dual solution which also has to sum up to the optimum with no slack space left to the primal.

The most challenging task is proving the dual constraints.
Basically, we have to prove $n^2$ inequalities on the dual variables.
If we consider the fact that the slack for cost incorporated in the final solution is zero, we can convert $n$ inequalities to equalities, however, because of the use of vector commitments they also have to be integrated into the same proof and cannot be done at the verifier.

To measure performance we implemented a version based on PySnark\footnote{\url{https://github.com/meilof/pysnark}} and QapTools\footnote{\url{https://github.com/Charterhouse/qaptools}} which is shown in Listing~\ref{lst:zkSNARK}.
It is a fully privacy-preserving version with index and optimum also hidden (inside a commitment), and proves the optimum of the primal and the dual as well as all dual constraints in a single proof.
The achieved performance is shown in Table~\ref{tab:snarks}.
From the figures it can be seen that the proof computation is the only relevant factor but is still faster than the optimization process.
It should be noted that the proof computation is independent of the network latency and does not need any communication between MPC nodes except for a final reconstruction step.

\begin{lstlisting}[language=Python, caption=zkSNARK example for optimality proof.,label={lst:zkSNARK}]
@pysnark.runtime.snark
def is_optimum(C, ind, opt, u, v):
    n = len(C)
    res = 1

    # verify primal optimum is correct
    cost = 0
    for r, c in ind:
        cost += C[r.value][c.value]
    res *= opt == cost

    # verify optimum primal-dual
    res *= opt == (sum(u) + sum(v))

    # verify dual constraints
    for i in range(n):
        for j in range(n):
            z_ij = u[i] + v[j]
            res *= z_ij <= C[i][j]
    return res
\end{lstlisting}

\begin{table}[ht]
\centering
\begin{tabular}[]{@{}cccc@{}}
\toprule
n  & genprog (s) & prove (s) & verify (s)\\
\midrule
10 & 0.03 & 1.9 & 0.07\\
20 & 0.10 & 3.5 & 0.07\\
30 & 0.22 & 5.7 & 0.07\\
40 & 0.39 & 9.4 & 0.11\\
50 & 0.61 & 18  & 0.21\\
60 & 0.89 & 21  & 0.21\\
70 & 1.23 & 37  & 0.40\\
80 & 1.71 & 42  & 0.42\\
90 & 2.15 & 65  & 0.40\\
100& 2.45 & 72  & 0.75\\
\bottomrule
\end{tabular}
\caption{Runtime in seconds for public verifiability of MPC based LSAP solving with adaptive zkSNARKs. The table shows results for different problem sizes $n$.}%
\label{tab:snarks}
\end{table}

\subsection{NIZK Without CRS}
Despite their practicality, zkSNARKs require a common reference string (CRS), which needs to be computed in a setup phase.
Although this CRS can also be generated in distributed way (e.g., in an MPC ceremony) in order to ensure that no entity knows, e.g., any trapdoor information of the CRS, it is sometimes undesirable to require a setup phase, in particular as the purpose of the NIZK is to protect against malicious MPC nodes, and thus an independent MPC network would be required for the MPC ceremony.

To also support a method without a CRS we leverage Bulletproofs \cite{DBLP:conf/sp/BunzBBPWM18}, which were designed to support efficient range proofs, the most demanding step when proving (\ref{eqn:dual-lsap}).
When batching $u$ interval proofs for intervals of bitlength $v$, the resulting proof size is only $2(\log_2(u) + \log_2(v)) + 4$ group elements plus 5 $\mathbb{Z}_p$ elements.

\begin{table}[ht]
\centering
\begin{tabular}[]{@{}ccccc@{}}
\toprule
$n$ & prove & verify & fast   & proof\\
          &       &        & verify & size\\
          & (s)   & (s)    & (s)    & (elem.)\\
\midrule
4   & 0.26   & 0.14   & 0.1    & 20\tabularnewline
5   & 0.25   & 0.27   & 0.2    & 22\tabularnewline
8   & 1.04   & 0.54   & 0.4    & 24\tabularnewline
11  & 2.07   & 1.09   & 0.81   & 26\tabularnewline
16  & 4.14   & 2.17   & 1.61   & 28\tabularnewline
22  & 8.28   & 4.33   & 3.22   & 30\tabularnewline
32  & 16.5  & 8.65   & 6.45   & 32\tabularnewline
45  & 33.0  & 17.2  & 12.8  & 34\tabularnewline
64  & 66.2  & 34.5  & 25.7  & 36\tabularnewline
90  & 132 & 68.8  & 51.5   & 38\tabularnewline
128 & 267 & 138 & 102 & 40\tabularnewline
\bottomrule
\end{tabular}
\caption{Runtime and size for optimality proofs based on Bulletproofs. Results for different problem sizes $n$ are shown.}%
\label{tab:bulletproofs}
\end{table}

Table~\ref{tab:bulletproofs} shows proof generation and verification times for different batch sizes and problem sizes.
The major issue at the moment are the size of the input commitments, because vector commitments are not supported and for each witness a dedicated Pedersen commitment is needed, which results in $n^2 + 2n$ commitments for the weight matrix and the dual variables.

The benchmark results only consider the computational intensive part of range proof processing, however, because the additional comparisons can be directly done at the verifier in parallel, they are good estimates for overall performance.
A fully integrated implementation also supporting MPC is currently under development.


%% file: sections/conclusion.tex
\section{Conclusion}%
\label{sec:conclusion}

Based on the results in our work it is easy to estimate the overall performance for solving the linear assignment problem in a privacy preserving but verifiable manner.
The times for MPC based optimization can be summed with the corresponding prove times for the given problem size to get an overall time.
Additionally the verification times is only done by the results parties.
Moreover, in some use cases the verifiability part could be done offline to further speed up overall application performance.

In this work we did a deep dive into solving the LSAP with MPC.
From our experience, it was not possible to arrive at this result without actually implementing the different solutions.
We were able to improve by a factor of 50 compared to the existing simplex based approach and also showed that shortest augmenting path solutions are the best also in the MPC setting.

On top of privacy, we also showed that efficient prove generation is possible by selecting most appropriate NIZK frameworks.
Compared to existing approach based on Schnorr proofs the usage of modern NIZK techniques showed major improvements and practical relevance.

%% file: main.bbl
\begin{thebibliography}{}

\bibitem[Akg{\"{u}}l, 1992]{Akgul1992}
Akg{\"{u}}l, M. (1992).
\newblock {The Linear Assignment Problem}.
\newblock In Akg{\"{u}}l, M., Hamacher, H.~W., and T{\"{u}}fek{\c{c}}i, S.,
  editors, {\em Combinatorial Optimization}, pages 85--122. Springer.

\bibitem[Aly and Cleemput, 2017]{AC2017}
Aly, A. and Cleemput, S. (2017).
\newblock {An Improved Protocol for Securely Solving the Shortest Path Problem
  and its Application to Combinatorial Auctions}.
\newblock Cryptology ePrint Archive, Report 2017/971.

\bibitem[Ausubel and Milgrom, 2002]{AusubelM02}
Ausubel, L. and Milgrom, P. (2002).
\newblock Ascending auctions with package bidding.
\newblock {\em Frontiers of Theoretical Economics}, 1(1):Article 1.

\bibitem[Baum et~al., 2014]{DBLP:conf/scn/BaumDO14}
Baum, C., Damg{\aa}rd, I., and Orlandi, C. (2014).
\newblock Publicly auditable secure multi-party computation.
\newblock In Abdalla, M. and Prisco, R.~D., editors, {\em {SCN} 2014}, volume
  8642 of {\em LNCS}, pages 175--196. Springer.

\bibitem[Bertsekas, 1979]{Bertsekas1979}
Bertsekas, D.~P. (1979).
\newblock {A distributed algorithm for the assignment problem}.
\newblock {\em Lab. for Information and Decision Systems Working Paper, MIT}.

\bibitem[Bertsekas, 1992]{Bertsekas1992}
Bertsekas, D.~P. (1992).
\newblock {Auction algorithms for network flow problems: A tutorial
  introduction}.
\newblock {\em Computational Optimization and Applications}, 1(1):7--66.

\bibitem[Bertsekas, 1998]{Bertsekas1998}
Bertsekas, D.~P. (1998).
\newblock {Network optimization : continuous and discrete models}.

\bibitem[Bertsekas, 2009]{Bertsekas2009}
Bertsekas, D.~P. (2009).
\newblock {Auction Algorithms}.
\newblock In {\em Encyclopedia of Optimization}.

\bibitem[Blum et~al., 1988]{DBLP:conf/stoc/BlumFM88}
Blum, M., Feldman, P., and Micali, S. (1988).
\newblock Non-interactive zero-knowledge and its applications (extended
  abstract).
\newblock In Simon, J., editor, {\em STOC 1988}, pages 103--112. {ACM}.

\bibitem[Boffey and Bertsekas, 1994]{Boffey1994}
Boffey, T.~B. and Bertsekas, D.~P. (1994).
\newblock {Linear Network Optimization: Algorithms and Codes.}
\newblock {\em The Journal of the Operational Research Society}, 45(4).

\bibitem[B{\"{u}}nz et~al., 2018]{DBLP:conf/sp/BunzBBPWM18}
B{\"{u}}nz, B., Bootle, J., Boneh, D., Poelstra, A., Wuille, P., and Maxwell,
  G. (2018).
\newblock Bulletproofs: Short proofs for confidential transactions and more.
\newblock In {\em {S\& P} 2018}, pages 315--334. {IEEE} Computer Society.

\bibitem[Burkard and {\c{C}}ela, 1999]{Burkard1999}
Burkard, R.~E. and {\c{C}}ela, E. (1999).
\newblock {\em {Linear Assignment Problems and Extensions}}, pages 75--149.
\newblock Springer.

\bibitem[Damg{\aa}rd et~al., 2017]{DDN+15}
Damg{\aa}rd, I., Damg{\aa}rd, K., Nielsen, K., Nordholt, P.~S., and Toft, T.
  (2017).
\newblock {Confidential Benchmarking Based on Multiparty Computation}.
\newblock In Grossklags, J. and Preneel, B., editors, {\em Financial
  Cryptography and Data Security}, pages 169--187, Berlin, Heidelberg. Springer
  Berlin Heidelberg.

\bibitem[de~Hoogh et~al., 2016]{HSM2016}
de~Hoogh, S., Schoenmakers, B., and Veeningen, M. (2016).
\newblock {Certificate Validation in Secure Computation and Its Use in
  Verifiable Linear Programming}.
\newblock In Pointcheval, D., Nitaj, A., and Rachidi, T., editors, {\em
  AFRICACRYPT 2016}, pages 265--284. Springer.

\bibitem[Dell'Amico and Toth, 2000]{DT00}
Dell'Amico, M. and Toth, P. (2000).
\newblock {Algorithms and codes for dense assignment problems: the state of the
  art}.
\newblock {\em Discrete Applied Mathematics}, 100(1):17--48.

\bibitem[{DFLEX}, 2014]{udpp}
{DFLEX} (2014).
\newblock {Demonstration Report (D2)}.
\newblock SESAR Joint Untertaking project deliverable.

\bibitem[Doerner et~al., 2016]{DBLP:conf/ccs/DoernerES16}
Doerner, J., Evans, D., and Shelat, A. (2016).
\newblock Secure stable matching at scale.
\newblock In Weippl, E.~R., Katzenbeisser, S., Kruegel, C., Myers, A.~C., and
  Halevi, S., editors, {\em {ACM} CCS 2016}, pages 1602--1613. {ACM}.

\bibitem[Franklin et~al., 2007]{DBLP:conf/ctrsa/FranklinGM07}
Franklin, M.~K., Gondree, M.~A., and Mohassel, P. (2007).
\newblock Improved efficiency for private stable matching.
\newblock In Abe, M., editor, {\em {CT-RSA} 2007}, volume 4377 of {\em LNCS},
  pages 163--177. Springer.

\bibitem[Gale and Shapley, 2013]{journals/tamm/GaleS13}
Gale, D. and Shapley, L.~S. (2013).
\newblock College admissions and the stability of marriage.
\newblock {\em The American Mathematical Monthly}, 120(5):386--391.

\bibitem[Golle, 2006]{DBLP:conf/fc/Golle06}
Golle, P. (2006).
\newblock A private stable matching algorithm.
\newblock In Crescenzo, G.~D. and Rubin, A.~D., editors, {\em {FC} 2006},
  volume 4107 of {\em LNCS}, pages 65--80. Springer.

\bibitem[Hong et~al., 2016]{DBLP:journals/corr/HongVRL16}
Hong, Y., Vaidya, J., Rizzo, N., and Liu, Q. (2016).
\newblock Privacy preserving linear programming.
\newblock {\em CoRR}, abs/1610.02339.

\bibitem[Jonker and Volgenant, 1987]{Jonker1987}
Jonker, R. and Volgenant, A. (1987).
\newblock {A shortest augmenting path algorithm for dense and sparse linear
  assignment problems}.
\newblock {\em Computing}, 38(4):325--340.

\bibitem[Kuhn, 1955]{Kuhn1955}
Kuhn, H.~W. (1955).
\newblock {The Hungarian method for the assignment problem}.
\newblock {\em Naval Research Logistics Quarterly}, 2(1-2).

\bibitem[Kuhn, 1956]{Kuhn1956}
Kuhn, H.~W. (1956).
\newblock {Variants of the hungarian method for assignment problems}.
\newblock {\em Naval Research Logistics Quarterly}, 3(4).

\bibitem[Li and Atallah, 2006]{DBLP:conf/colcom/0001A06}
Li, J. and Atallah, M.~J. (2006).
\newblock Secure and private collaborative linear programming.
\newblock In Blanzieri, E. and Zhang, T., editors, {\em CollaborateCom 2006}.
  {IEEE} Computer Society / {ICST}.

\bibitem[Loruenser et~al., 2022]{LRW22}
Loruenser, T., Rainer, B., and Wohner, F. (2022).
\newblock {Towards a Performance Model for Byzantine Fault Tolerant Services}.
\newblock In {\em Proceedings of the 12th International Conference on Cloud
  Computing and Services Science - CLOSER,}, pages 178--189. INSTICC,
  SciTePress.

\bibitem[Lor{\"{u}}nser et~al., 2021]{LSG21}
Lor{\"{u}}nser, T., Sch{\"{u}}tz, C.~G., and Gringinger, E. (2021).
\newblock {SlotMachine - A Privacy-preserving Marketplace for Slot Management}.
\newblock {\em ERCIM News}, 2021(126).

\bibitem[Lor{\"{u}}nser and Wohner, 2020]{LW20}
Lor{\"{u}}nser, T. and Wohner, F. (2020).
\newblock {Performance Comparison of Two Generic MPC-frameworks with Symmetric
  Ciphers}.
\newblock In {\em Proceedings of the 17th International Joint Conference on
  e-Business and Telecommunications}, pages 587--594. SCITEPRESS - Science and
  Technology Publications.

\bibitem[Lor{\"{u}}nser et~al., 2022]{DBLP:conf/icissp/LorunserWK22}
Lor{\"{u}}nser, T., Wohner, F., and Krenn, S. (2022).
\newblock A privacy-preserving auction platform with public verifiability for
  smart manufacturing.
\newblock In Mori, P., Lenzini, G., and Furnell, S., editors, {\em {ICISSP}
  2022}, pages 637--647. {SCITEPRESS}.

\bibitem[Moldovanu, 2012]{Moldovanu12}
Moldovanu, B. (2012).
\newblock Auction theory and applications.
\newblock {\em The Bonn Journal of Economics}, 1(1):53--64.

\bibitem[Munkres, 1957]{Munkres1957}
Munkres, J. (1957).
\newblock {Algorithms for the Assignment and Transportation Problems}.
\newblock {\em Journal of the Society for Industrial and Applied Mathematics},
  5(1).

\bibitem[Ramshaw and Tarjan, 2012]{doubling-technique}
Ramshaw, L. and Tarjan, R.~E. (2012).
\newblock {On Minimum-Cost Assignments in Unbalanced Bipartite Graphs}.
\newblock HP Laboratories Technical Report, HPL-2012-40R1.

\bibitem[Riazi et~al., 2017]{DBLP:journals/popets/RiaziSSSK17}
Riazi, M.~S., Songhori, E.~M., Sadeghi, A., Schneider, T., and Koushanfar, F.
  (2017).
\newblock Toward practical secure stable matching.
\newblock {\em PoPETs}, 2017(1):62--78.

\bibitem[Schoenmakers and Veeningen, 2015]{SM2015}
Schoenmakers, B. and Veeningen, M. (2015).
\newblock {Universally Verifiable Multiparty Computation from Threshold
  Homomorphic Cryptosystems}.
\newblock In Malkin, T., Kolesnikov, V., Lewko, A.~B., and Polychronakis, M.,
  editors, {\em ACNS 2015}, pages 3--22. Springer.

\bibitem[Schuetz et~al., 2021]{SGPL21}
Schuetz, C.~G., Gringinger, E., Pilon, N., and Lor{\"{u}}nser, T. (2021).
\newblock {A Privacy-Preserving Marketplace for Air Traffic Flow Management
  Slot Configuration}.
\newblock In {\em 2021 IEEE/AIAA 40th Digital Avionics Systems Conference
  (DASC)}, pages 1--9.

\bibitem[Sunderam and Parkes, 2003]{DBLP:conf/sigecom/SunderamP03}
Sunderam, A.~V. and Parkes, D.~C. (2003).
\newblock Preference elicitation in proxied multiattribute auctions.
\newblock In Menasc{\'{e}}, D.~A. and Nisan, N., editors, {\em Electronic
  Commerce - EC 2003}, pages 214--215. {ACM}.

\bibitem[Toft, 2009]{Toft09}
Toft, T. (2009).
\newblock Solving linear programs using multiparty computation.
\newblock In Dingledine, R. and Golle, P., editors, {\em {FC} 2009}, volume
  5628 of {\em LNCS}, pages 90--107. Springer.

\bibitem[Vaidya, 2009]{DBLP:conf/sac/Vaidya09}
Vaidya, J. (2009).
\newblock Privacy-preserving linear programming.
\newblock In Shin, S.~Y. and Ossowski, S., editors, {\em SAC 2009}, pages
  2002--2007. {ACM}.

\bibitem[Veeningen, 2017]{DBLP:conf/africacrypt/Veeningen17}
Veeningen, M. (2017).
\newblock Pinocchio-based adaptive zk-snarks and secure/correct adaptive
  function evaluation.
\newblock In Joye, M. and Nitaj, A., editors, {\em {AFRICACRYPT} 2017}, volume
  10239 of {\em LNCS}, pages 21--39.

\bibitem[W{\"{u}}ller et~al., 2017]{Wuller2017a}
W{\"{u}}ller, S., Vu, M., Meyer, U., and Wetzel, S. (2017).
\newblock {Using Secure Graph Algorithms for the Privacy-Preserving
  Identification of Optimal Bartering Opportunities}.
\newblock In {\em WPES 2017}, pages 123--132. ACM.

\bibitem[Yao, 1986]{DBLP:conf/focs/Yao86}
Yao, A.~C. (1986).
\newblock How to generate and exchange secrets (extended abstract).
\newblock In {\em FOCS 1986}, pages 162--167. {IEEE} Computer Society.

\end{thebibliography}
